\documentclass[sigconf,nonacm]{acmart}

\setcopyright{none}
\renewcommand\footnotetextcopyrightpermission[1]{}
\usepackage{glossaries}
\usepackage{tikz} 
\usetikzlibrary{arrows.meta,positioning,fit,backgrounds}
\usepackage{cleveref}
\usepackage{booktabs}
\usepackage{tabularx}
\usepackage{makecell}
\usepackage{placeins}

\begin{document}
\newacronym{vqe}{VQE}{Variational Quantum Eigensolver}
\newacronym{vqa}{VQA}{Variational Quantum Algorithm}
\newacronym{vqas}{VQAs}{Variational Quantum Algorithms}
\newacronym{shacl}{SHACL}{Shapes Constraint Language}
\newacronym{sparql}{SPARQL}{SPARQL Protocol and RDF Query Language}
\newacronym{qml}{QML}{Quantum Machine Learning}
\newacronym{owl}{OWL}{Web Ontology Language}
\newacronym{qaoa}{QAOA}{Quantum Approximate Optimization Algorithm}
\newacronym{rdf}{RDF}{Resource Description Framework}
\newacronym{hpc}{HPC}{High Performance Computing}

\title{A Semantic Framework for Reproducible Variational Quantum Algorithm Execution Records}

\author{Silvie Ill\'{e}sov\'{a}}
\affiliation{%
  \institution{Gran Sasso Science Institute}
  \city{L'Aquila}
  \country{Italy}
}
\email{silvie.illesova@gssi.it}

\author{Martin Beseda}
\affiliation{%
  \institution{Department of Information Engineering, Computer Science and Mathematics, University of L'Aquila}
  \city{L'Aquila}
  \country{Italy}
}
\email{martin.beseda@univaq.it}

\begin{abstract} 
Variational quantum algorithms are hybrid quantum--classical workflows whose results depend on many interacting choices, including the ansatz, Hamiltonian, optimizer, backend, shot count, noise model, mitigation method, random seed, stopping criteria, and software versions. In current practice, this information is often scattered across code, configuration files, logs, backend metadata, and paper descriptions, making executions difficult to reproduce, compare, debug, and reuse. This paper proposes an ontology-supported framework for representing \gls{vqa} execution records as structured and machine-readable software-engineering artifacts. The framework defines a \gls{owl} ontology for modeling the main entities involved in \gls{vqa} experimentation, including algorithms, circuits, ansatzes, Hamiltonians, optimizers, backends, noise models, mitigation techniques, execution steps, software environments, measurement outcomes, and results. It further combines the ontology with \gls{shacl} constraints for validating completeness and consistency, and \gls{sparql} competency queries for retrieving reproducibility-relevant information. We demonstrate the approach using \gls{vqe} execution records, including valid record and intentionally incomplete or inconsistent examples. The results show that the framework can represent complete \gls{vqa} execution contexts, detect missing or malformed metadata, and support query-based inspection of information needed for reproducible quantum software experimentation. 
\end{abstract}


\maketitle

\section{Introduction}

\gls{vqas} constitute a prominent class of near-term quantum algorithms \cite{cerezo2021variational,qi2024variational,xu2024quantum}, especially in the Noisy Intermediate-Scale Era \cite{novak2025optimization,novak2025reliable,bezdek2025classical}. They integrate parameterized quantum circuits with classical optimization procedures and are widely applied in multiple different fields, including, but not limited to quantum chemistry \cite{delgado2021variational,singh2023benchmarking,bauer2020quantum,illesova2025statistical,ciaramelletti2025detecting}, combinatorial optimization \cite{amaro2022filtering,wu2026resource,li2026variational}, finance \cite{herman2023quantum,wang2025variational,thi2024variational}, cybersecurity \cite{rahman2024fine,moll2025case,rahgozar2025quantum}, and \gls{qml} \cite{tacchino2021variational,chen2020variational,livingston2025application,novak2025predicting,novak2026quantum} or hybrid machine learning \cite{illesova2025complementarity,illesova2025classical}. From a software engineering perspective, a \gls{vqa} extends beyond an algorithmic concept or circuit template \cite{njoku2025quantum,mandal2025quantum,scheerer2023experiences}. It represents a complex hybrid software workflow whose behavior depends on the interplay among quantum circuits, classical optimizers, numerical parameters, execution platforms, stochastic sampling, noise models, software libraries, and hardware or simulator backends.

Taking this all into account, engineering \gls{vqa} experimentation in a reliable manner is challenging. A single execution may depend on factors such as ansatz structure, Hamiltonian representation, optimizer configuration, shot count, backend calibration, noise model, mitigation technique, random seed, stopping criterion, software version, and measurement strategy \cite{weder2023provenance,buonaiuto2024effects,gokhale2019partial,nakaji2023measurement,illesova2025qmetric}. These choices are intertwined as, for example, changing the backend may necessitate a different noise model or altering the shot count can influence optimizer behavior. Therefore, the outcome of a \gls{vqa} execution cannot be fully understood based solely on the final energy value or accuracy metric. Interpretation requires consideration of the entire execution context.

In current practice, this execution context is often distributed across source code, configuration files, log files, backend metadata, and textual descriptions in research papers. Such fragmentation creates several software engineering problems. It makes experiments harder to reproduce, because essential execution details may be missing or implicit. It makes debugging harder because failures or unexpected convergence behavior cannot easily be traced back to specific configuration choices. It makes comparison harder because two reported \gls{vqa} results may differ in hidden assumptions such as optimizer settings, shot allocation, noise simulation, or software versions. It also limits reuse, since execution records are rarely represented in a form that can be shared, validated, queried, and integrated with other tools.

Existing quantum software frameworks \cite{javadi2024quantum,Cirq_Developers_2025,bergholm2018pennylane} provide extensive support for building circuits, running algorithms, simulating noise, and collecting results. However, they do not provide a standard semantic representation of  \gls{vqa} execution records. The relationships between the algorithm, ansatz, Hamiltonian, optimizer, backend, noise model, mitigation method, execution steps, software environment, and result are usually not represented as first-class, queryable entities. As a result, important information about the software execution remains implicit or tool-specific.

To address this gap, we propose an ontology-supported framework for representing \gls{vqa} executions as structured and machine-readable metadata. The ontology models key concepts involved in \gls{vqa} experimentation, including algorithms, ansatzes, Hamiltonians, optimizers, backends, noise models, mitigation techniques, execution steps, software environments, parameter assignments, measurement outcomes, and results. By representing these concepts and their relationships explicitly, the framework supports a more systematic treatment of \gls{vqa} experiments as software engineering artifacts rather than isolated numerical runs.

The proposed framework also includes a validation and querying layer. \gls{shacl} \cite{pareti2021review,bogaerts2022shacl} constraints are used to check whether execution records contain the metadata required for completeness and consistency, while \gls{sparql} \cite{perez2009semantics} queries are used to retrieve information relevant to reproducibility, debugging, and comparison. This allows execution records to be inspected automatically, for example, to identify which backend, optimizer, shot count, software version, or noise model was used in a given run, or to detect records where required information is missing.

We demonstrate the framework using \gls{vqe} execution records. The case study includes valid record as well as intentionally incomplete or inconsistent examples, showing how the ontology captures the structure of \gls{vqe} workflows, how validation can detect metadata problems, and how queries can support reproducibility-oriented analysis. 

The contributions of this paper are:
\begin{itemize}
\item an \gls{owl} ontology for representing \gls{vqa} execution metadata, including algorithmic, circuit-level, backend, noise, optimization, software-environment, \gls{hpc}-resource, and result information;
\item a \gls{shacl} validation layer for detecting incomplete, malformed, or inconsistent \gls{vqa} execution records;
\item a set of \gls{sparql} competency queries for retrieving reproducibility-relevant execution metadata;
\item a \gls{vqe} case study with valid and intentionally invalid execution records, demonstrating representation, validation, and querying.
\end{itemize}

\section{Background and Motivation}

This section introduces the technical context of the proposed framework. It summarizes \gls{vqa} workflows, discusses reproducibility challenges, and motivates the need for structured execution records.

\subsection{Variational Quantum Algorithms}
\begin{figure}[ht!]
\centering
\resizebox{\columnwidth}{!}{%
\begin{tikzpicture}[
    box/.style={
        draw,
        rounded corners=2pt,
        align=center,
        minimum width=4.4cm,
        minimum height=0.85cm,
        font=\small,
        fill=white
    },
    usual/.style={
        draw,
        rounded corners=2pt,
        align=center,
        minimum width=4.4cm,
        minimum height=0.85cm,
        font=\scriptsize,
        fill=orange!12
    },
    desired/.style={
        draw,
        dashed,
        rounded corners=2pt,
        align=center,
        minimum width=4.9cm,
        minimum height=0.95cm,
        font=\scriptsize,
        fill=green!12
    },
    arrow/.style={
        -{Latex[length=2mm]},
        thick
    }
]

\node[box] (setup) at (0,0) {
    VQA experiment setup\\
    \scriptsize problem, ansatz, backend, optimizer, shots, seed
};

\node[box] (circuit) at (0,-1.35) {
    Parameterized quantum circuit\\
    $U(\boldsymbol{\theta})$
};

\node[box] (execution) at (0,-2.70) {
    Circuit translation and transpilation\\
    with respect to the configured backend
};

\node[box] (measurement) at (0,-4.05) {
    Execution, measurement, expectation\\
    value estimation and mitigation
};

\node[box] (cost) at (0,-5.40) {
    Cost-function evaluation\\
    $C(\boldsymbol{\theta})$
};

\node[box] (check) at (0,-6.75) {
    Convergence criteria satisfied?
};

\node[box] (result) at (0,-8.10) {
    Final parameters and result
};

\node[box] (optimizer) at (5.2,-6.75) {
    Classical optimizer\\
    updates parameters
};

\node[usual] (usual) at (0,-9.50) {
    Usual output\\
    final energy, final parameters
};

\node[desired] (desired) at (5.2,-9.50) {
    Desired output\\
    machine-readable execution record:\\
    setup, software versions, noise model,\\
    mitigation, results
};

\begin{scope}[on background layer]
\node[
    fill=blue!8,
    draw=blue!35,
    rounded corners=6pt,
    fit=(circuit)(execution)(measurement)(cost)(check)(optimizer),
    inner xsep=0.35cm,
    inner ysep=0.35cm
] (loopbg) {};
\end{scope}

\node[
    font=\small\bfseries,
    text=blue!55!black,
    fill=blue!8,
    inner sep=1.5pt,
    anchor=north east,
    xshift=-4pt,
    yshift=-4pt
] at (loopbg.north east) {Hybrid execution loop};

\draw[arrow] (setup) -- (circuit);
\draw[arrow] (circuit) -- node[right, font=\scriptsize] {$\boldsymbol{\theta}$} (execution);
\draw[arrow] (execution) -- (measurement);
\draw[arrow] (measurement) -- node[right, font=\scriptsize] {expectation values} (cost);
\draw[arrow] (cost) -- (check);
\draw[arrow] (check) -- node[right, font=\scriptsize] {yes} (result);
\draw[arrow] (result) -- (usual);

\draw[arrow] (check.east) -- node[above, font=\scriptsize] {no} (optimizer.west);

\draw[arrow]
    (optimizer.north)
    -- ++(0,5.0)
    -- node[above, font=\scriptsize] {updated $\boldsymbol{\theta}$} (circuit.east);

\draw[arrow] (result) -- (usual);

\draw[arrow, dashed]
    (usual.south)
    -- ++(0,-0.45)
    -| node[pos=0.25, below, font=\scriptsize] {extended record}
    (desired.south);

\draw[arrow, dashed]
    (setup.east)
    -- ++(6.4,0)
    |- (desired.east);

\draw[arrow, dashed]
    (setup.east)
    -- ++(6.4,0)
    |- (desired.east);

\end{tikzpicture}%
}
\caption{General workflow of a variational quantum algorithm execution. }
\Description{A flow diagram of a variational quantum algorithm execution. The workflow starts with the \gls{vqa} experiment setup, followed by a parameterized quantum circuit, circuit translation and transpilation, execution and measurement, cost-function evaluation, convergence checking, and final parameters and result. If the convergence criterion is not satisfied, a classical optimizer updates the parameters and the process returns to the circuit. The usual output is the final energy and final parameters, while the desired software-engineering output is a machine-readable execution record containing setup information, software versions, noise model, mitigation settings, and results.} 
\label{fig:vqa-workflow}
\end{figure}
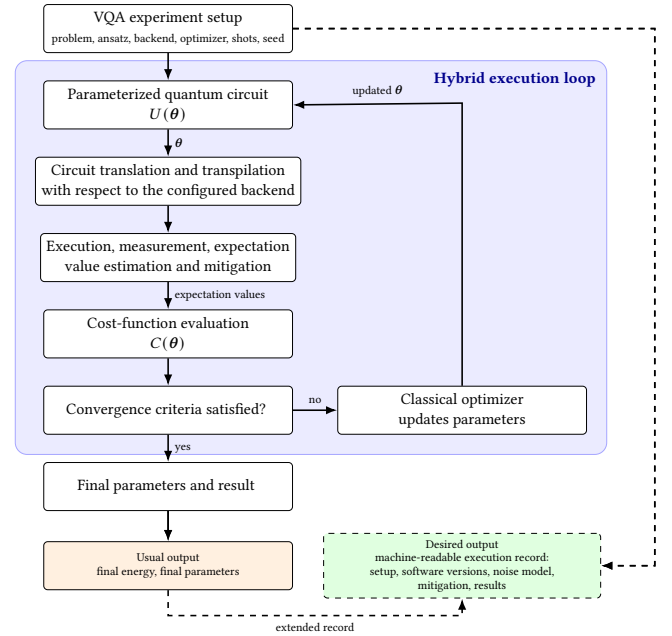

\gls{vqas} are hybrid quantum--classical algorithms designed for near-term quantum devices. They combine a parameterized quantum circuit with a classical optimizer. The quantum part prepares trial states and estimates expectation values, while the classical part updates the circuit parameters. The generalized execution workflow is displayed in \Cref{fig:vqa-workflow}. Important examples include the \gls{vqe} \cite{tilly2022variational,kandala2017hardware,liu2019variational,rajamani2025equi, beseda2024state,illesova2025transformation}, the \gls{qaoa} \cite{zhou2020quantum,guerreschi2019qaoa,trovato2025preliminary}, and \gls{qml} \cite{illesova2025importance,gupta2022how}. In this work, we focus on \gls{vqa} execution records, using \gls{vqe} as the main motivating example.

\subsection{Reproducibility and Metadata Challenges}

\gls{vqa} results depend on many experimental and implementation choices, including the ansatz \cite{wu2021towards,qin2023review,choquette2021quantum}, Hamiltonian, optimizer, backend, shot count, noise model, mitigation method \cite{ravi2022vaqem,botelho2022error,barron2020measurement}, random seed, stopping criteria, and software versions. In practice, these details are often distributed across source code, configuration files, logs, backend metadata, and paper descriptions. This makes it difficult to reproduce reported results, compare different executions, and reuse experimental artifacts \cite{senapati2023towards,qi2024variational}. From a quantum software engineering perspective, this is a metadata and traceability problem as reliable \gls{vqa} experimentation requires execution records that are complete, structured, and machine-checkable.

\subsection{Motivating VQE Execution Scenario}

Consider a \gls{vqe} execution for estimating the ground-state energy of a molecular Hamiltonian. Such an execution involves a parameterized ansatz circuit, a Pauli-decomposed Hamiltonian, a classical optimizer, and a selected backend or simulator. To reproduce the run, additional information such as the shot count, random seed, convergence criterion, noise model, and software environment is also required. If these details are represented only informally, checking completeness and consistency requires manual inspection. This motivates an ontology-based representation of \gls{vqa} executions, combined with \gls{shacl} validation and \gls{sparql} queries for reproducibility-oriented analysis.

\section{Ontology-Supported Representation of VQA Executions}

\begin{table*}[t]
\centering
\small
\setlength{\tabcolsep}{4pt}
\renewcommand{\arraystretch}{0.95}
\caption{Classes in the proposed \gls{vqa} ontology.}
\label{tab:ontology-classes}
\begin{tabularx}{\textwidth}{@{}
    >{\raggedright\arraybackslash}p{0.18\textwidth}
    >{\raggedright\arraybackslash}X
@{}}
\toprule
\textbf{Module} & \textbf{Ontology classes} \\
\midrule

Execution and workflow &
\texttt{Execution}, \texttt{HybridExecution}, \texttt{ExecutionStep}, \texttt{QuantumAlgorithmStep}, \texttt{ClassicalAlgorithmStep}, \texttt{AlgorithmStep}, \texttt{ProblemSolving}, \texttt{VQATask} \\[0.8mm]

Algorithm &
\texttt{Algorithm}, \texttt{HybridAlgorithm}, \texttt{ClassicalAlgorithm}, \texttt{VariationalQuantumAlgorithm}, \texttt{VQE}, \texttt{QAOA}, \texttt{SAOOVQE} \\[0.8mm]

Problem and objective &
\texttt{ProblemHamiltonian}, \texttt{CostFunction}, \texttt{Observable}, \texttt{PauliObservable}, \texttt{HamiltonianTerm}, \texttt{PauliTerm} \\[0.8mm]

Circuit and parameters &
\texttt{QuantumCircuit}, \texttt{FixedQuantumCircuit}, \texttt{ParameterizedQuantumCircuit}, \texttt{Ansatz}, \texttt{CircuitParameter}, \texttt{ParameterAssignment}, \texttt{ParameterDomain}, \texttt{AdaptiveAnsatz}, \texttt{HardwareEfficientAnsatz}, \texttt{ProblemInspiredAnsatz}, \texttt{ChemicallyMotivatedAnsatz}, \texttt{UCCAnsatz}, \texttt{UCCSDAnsatz} \\[0.8mm]

Optimization &
\texttt{ClassicalOptimizer}, \texttt{ConvergenceCriterion}, \texttt{OptimizationTrace}, \texttt{IterativeSolver}, \texttt{DirectSolver} \\[0.8mm]

Backend and resources &
\texttt{QuantumBackend}, \texttt{ComputationalResource}, \texttt{BackendCalibration}, \texttt{StatevectorSimulator}, \texttt{NoisySimulator}, \texttt{QuantumHardwareBackend}, \texttt{QuantumProcessingUnit}, \texttt{ClassicalCluster} \\[0.8mm]

Noise and mitigation &
\texttt{NoiseModel}, \texttt{ErrorMitigationTechnique}, \texttt{ReadoutNoise}, \texttt{DepolarizingNoise}, \texttt{AmplitudeDampingNoise}, \texttt{CalibrationDerivedNoise}, \texttt{ZNE}, \texttt{M3}, \texttt{MEM}, \texttt{TREX} \\[0.8mm]

Measurement and results &
\texttt{MeasurementStrategy}, \texttt{GroupedMeasurementStrategy}, \texttt{MeasurementOutcome}, \texttt{ExecutionResult} \\[0.8mm]

Software and execution context &
\texttt{SoftwareEnvironment}, \texttt{SoftwarePackage}, \texttt{ExecutionContextRecord} \\[0.8mm]

Auxiliary execution concepts &
\texttt{InitializeCircuit}, \texttt{AssembleCircuits}, \texttt{MeasureCircuit}, \texttt{DataPreprocessing}, \texttt{DataPostprocessing}, \texttt{AdaptiveMeasurementAlgorithm}, \texttt{HybridQuantumErrorCorrection} \\

\bottomrule
\end{tabularx}
\end{table*}
We propose an \gls{owl} ontology for representing \gls{vqa} executions as structured and machine-readable software-engineering artifacts. The ontology is intended to capture not only the final numerical output of a \gls{vqa} run, but also the execution context needed to reproduce, validate, compare, and reuse the result. In particular, it represents the main entities involved in a \gls{vqa} execution, including the problem, ansatz, parameterized circuit, optimizer, backend, noise model, mitigation method, software environment, execution steps, and final result. \Cref{tab:ontology-classes} summarizes the main ontology classes grouped by conceptual module.

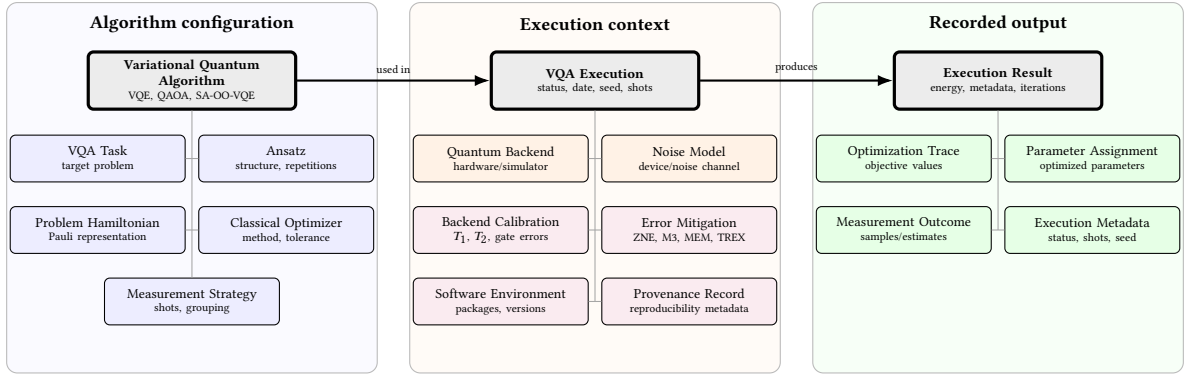
\begin{figure*}[t]
\centering
\begin{tikzpicture}[
    scale=0.86,
    transform shape,
    font=\scriptsize,
    box/.style={
        draw,
        rounded corners=2pt,
        align=center,
        text width=2.55cm,
        minimum height=0.72cm,
        inner sep=2pt
    },
    core/.style={
        box,
        very thick,
        fill=gray!15,
        font=\scriptsize\bfseries,
        text width=3.05cm,
        minimum height=0.88cm
    },
    algbox/.style={box, fill=blue!7},
    contextbox/.style={box, fill=orange!10},
    resultbox/.style={box, fill=green!10},
    metabox/.style={box, fill=purple!8},
    mainarr/.style={-{Latex[length=2mm]}, thick},
    assoc/.style={thin, gray!70},
    title/.style={font=\small\bfseries}
]

\draw[rounded corners=4pt, fill=blue!2, draw=gray!55]
    (-2.85,-2.55) rectangle (2.85,3.15);

\draw[rounded corners=4pt, fill=orange!3, draw=gray!55]
    (3.35,-2.55) rectangle (9.05,3.15);

\draw[rounded corners=4pt, fill=green!3, draw=gray!55]
    (9.55,-2.55) rectangle (15.25,3.15);

\node[title] at (0,2.82) {Algorithm configuration};
\node[title] at (6.20,2.82) {Execution context};
\node[title] at (12.40,2.82) {Recorded output};

\node[core] (vqa) at (0,1.95)
{Variational Quantum\\Algorithm\\[-1pt]
{\normalfont\tiny VQE, QAOA, SA-OO-VQE}};

\node[algbox] (task) at (-1.45,0.75)
{VQA Task\\[-1pt]
{\tiny target problem}};

\node[algbox] (ansatz) at (1.45,0.75)
{Ansatz\\[-1pt]
{\tiny structure, repetitions}};

\node[algbox] (ham) at (-1.45,-0.35)
{Problem Hamiltonian\\[-1pt]
{\tiny Pauli representation}};

\node[algbox] (opt) at (1.45,-0.35)
{Classical Optimizer\\[-1pt]
{\tiny method, tolerance}};

\node[algbox] (meas) at (0,-1.45)
{Measurement Strategy\\[-1pt]
{\tiny shots, grouping}};

\node[core] (exec) at (6.20,1.95)
{VQA Execution\\[-1pt]
{\normalfont\tiny status, date, seed, shots}};

\node[contextbox] (backend) at (4.75,0.75)
{Quantum Backend\\[-1pt]
{\tiny hardware/simulator}};

\node[contextbox] (noise) at (7.65,0.75)
{Noise Model\\[-1pt]
{\tiny device/noise channel}};

\node[metabox] (calib) at (4.75,-0.35)
{Backend Calibration\\[-1pt]
{\tiny $T_1$, $T_2$, gate errors}};

\node[metabox] (mit) at (7.65,-0.35)
{Error Mitigation\\[-1pt]
{\tiny ZNE, M3, MEM, TREX}};

\node[metabox] (soft) at (4.75,-1.45)
{Software Environment\\[-1pt]
{\tiny packages, versions}};

\node[metabox] (prov) at (7.65,-1.45)
{Provenance Record\\[-1pt]
{\tiny reproducibility metadata}};

\node[core] (res) at (12.40,1.95)
{Execution Result\\[-1pt]
{\normalfont\tiny energy, metadata, iterations}};

\node[resultbox] (trace) at (10.95,0.75)
{Optimization Trace\\[-1pt]
{\tiny objective values}};

\node[resultbox] (params) at (13.85,0.75)
{Parameter Assignment\\[-1pt]
{\tiny optimized parameters}};

\node[resultbox] (outcome) at (10.95,-0.35)
{Measurement Outcome\\[-1pt]
{\tiny samples/estimates}};

\node[resultbox] (status) at (13.85,-0.35)
{Execution Metadata\\[-1pt]
{\tiny status, shots, seed}};

\draw[assoc] (vqa.south) -- (0,1.20) -- (0,-1.08);

\draw[assoc] (0,0.75) -- (task.east);
\draw[assoc] (0,0.75) -- (ansatz.west);

\draw[assoc] (0,-0.35) -- (ham.east);
\draw[assoc] (0,-0.35) -- (opt.west);

\draw[assoc] (0,-1.08) -- (meas.north);

\draw[assoc] (exec.south) -- (6.20,1.20) -- (6.20,-1.45);

\draw[assoc] (6.20,0.75) -- (backend.east);
\draw[assoc] (6.20,0.75) -- (noise.west);

\draw[assoc] (6.20,-0.35) -- (calib.east);
\draw[assoc] (6.20,-0.35) -- (mit.west);

\draw[assoc] (6.20,-1.45) -- (soft.east);
\draw[assoc] (6.20,-1.45) -- (prov.west);

\draw[assoc] (res.south) -- (12.40,1.20) -- (12.40,-0.35);

\draw[assoc] (12.40,0.75) -- (trace.east);
\draw[assoc] (12.40,0.75) -- (params.west);

\draw[assoc] (12.40,-0.35) -- (outcome.east);
\draw[assoc] (12.40,-0.35) -- (status.west);

\draw[mainarr] (vqa.east) -- node[above,font=\tiny] {used in} (exec.west);
\draw[mainarr] (exec.east) -- node[above,font=\tiny] {produces} (res.west);

\end{tikzpicture}
\caption{Compact overview of the VQA ontology as an execution record. The ontology connects algorithm configuration, execution context, and recorded outputs required for validation, querying, and reproducibility.}
\label{fig:vqa-ontology-overview}
\end{figure*}

The current version of the ontology contains 71 \gls{owl} classes, 35 object properties, and 33 datatype properties. The overview is visualized in \Cref{fig:vqa-ontology-overview}. Its structure is organized around the central concept of a \gls{vqa} execution, which is connected to the algorithm being executed, the problem representation, the parameterized circuit, the classical optimizer, the selected backend, the software environment, and the produced result. The ontology also represents \gls{vqa}-specific metadata such as ansatz families, Pauli-decomposed Hamiltonians, circuit parameters, parameter assignments, measurement outcomes, backend calibration data, noise models, mitigation techniques, optimization traces, and provenance records. This organization allows a final numerical result to be linked explicitly to the software, hardware, numerical, and configuration choices that produced it.

The inclusion of computational-resource and software-environment concepts also allows the ontology to represent hybrid execution settings beyond the quantum backend itself. For example, a \gls{vqa} execution may involve quantum circuit evaluations performed on a simulator or quantum device, while classical optimization, preprocessing, postprocessing, logging, or orchestration are executed on local machines or \gls{hpc} clusters. In this way, the ontology captures the broader hybrid quantum--classical execution context rather than only the quantum-circuit execution.

The ontology is divided into conceptual modules corresponding to the main parts of a \gls{vqa} workflow. Algorithm-level classes describe \gls{vqa} methods and their variants, such as \gls{vqe} or \gls{qaoa}. Problem-level classes represent Hamiltonians, observables, objectives, and Pauli terms. Circuit-level classes describe ansatzes, parameterized circuits, circuit parameters, domains, and assignments. Optimization classes capture classical optimizers, stopping criteria, and optimization traces. Backend, noise, and mitigation classes describe simulators, quantum hardware backends, calibration information, noise models, and mitigation techniques. Finally, measurement, result, software, and provenance classes represent measurement outcomes, final results, software packages, execution environments, and metadata required for reproducibility.

A central aspect of the ontology is the explicit representation of the hybrid execution loop. A \gls{vqa} run is modeled as a sequence of quantum and classical steps in which a parameterized circuit is prepared, translated or transpiled, executed on a selected backend, measured, and used to estimate expectation values. These values are then consumed by a classical cost function and optimizer, which updates the parameters until a convergence criterion is satisfied. By representing these steps and their input and output artifacts explicitly, the ontology allows the final result to be traced back to the configuration choices and intermediate execution data that produced it.

\section{Validation and Querying}
The ontology representation described in the previous section is intended not only as a vocabulary for describing \gls{vqa} executions, but also as a basis for automatic inspection of execution records. For this reason, the framework includes two complementary mechanisms. First, \gls{shacl} constraints are used to check whether an RDF execution record contains the metadata required by the ontology and whether the values are internally consistent. Second, \gls{sparql} competency queries are used to retrieve information that is relevant for reproducibility-oriented analysis, such as the selected ansatz, optimizer, backend, Hamiltonian terms, software environment, and final result.

\subsection{SHACL-Based Validation} 
\gls{shacl} validation is used to express structural and consistency requirements for \gls{vqa} execution records. In the implemented workflow, the ontology and a \gls{rdf} execution record are loaded into an \gls{rdf} graph and validated against the \gls{shacl} shapes. The validation is performed with \gls{rdf} inference enabled, so that constraints can also apply through the class hierarchy defined in the ontology. The output consists of a conformance value and a validation report describing any detected violations. 

This step is important because many problems in \gls{vqa} reporting are metadata problems rather than errors in the final numerical value. For example, a run may be marked as completed without a valid result, a Hamiltonian term may contain an invalid Pauli string, or a backend may be described with an unsupported backend type. \gls{shacl} makes these requirements explicit and allows such problems to be detected automatically.  The valid \gls{vqe} record conforms to these constraints, while the intentionally invalid record produces 20 violations. These violations include datatype errors, invalid controlled-vocabulary values, invalid numerical ranges, malformed Pauli terms, missing result information, and inconsistencies between related metadata values.








\subsection{Validation Rules}

The validation rules check whether a \gls{vqa} execution record contains the information required for reproducibility-oriented analysis. They cover both structural requirements and value-level constraints. Structural checks verify that an execution is connected to the relevant algorithm, circuit, optimizer, backend, and result entities. Value-level checks verify that numerical and categorical metadata are well formed. For example, shot counts and iteration counts must be non-negative, parameter bounds must be valid, error rates must lie in admissible ranges, optimizer names must belong to the supported vocabulary, and Pauli strings must contain only valid Pauli operators. These rules make the expected metadata explicit and allow incomplete or inconsistent records to be detected automatically.

\subsection{Competency Questions}

In addition to validation, the framework is evaluated using competency questions expressed as \gls{sparql} queries over the \gls{rdf} execution records. Competency questions define the types of questions that the ontology should be able to answer and therefore provide a practical way to assess whether the modeled concepts and relationships are sufficient for the intended use case. In this work, the competency questions focus on reproducibility-oriented inspection of \gls{vqa} execution records. The questions are summarized in \Cref{tab:sparql-queries}.

\begin{table*}[t]
\centering
\small
\caption{Competency questions used for inspecting \gls{vqa} execution records.}
\label{tab:sparql-queries}
\begin{tabular}{p{0.07\textwidth}p{0.45\textwidth}p{0.40\textwidth}}
\toprule
\textbf{ID} & \textbf{Competency question} & \textbf{Retrieved or detected information} \\
\midrule
CQ1 & Which algorithmic components define a \gls{vqe} execution? & Algorithm instance, ansatz, classical optimizer, and problem Hamiltonian. \\
CQ2 & Which executions were completed, and what results did they produce? & Execution identifier, backend, final energy, energy uncertainty, and optimizer iteration count. \\
CQ3 & How is the problem Hamiltonian represented in terms of Pauli operators? & Hamiltonian identifier, Pauli strings, and corresponding coefficients. \\
CQ4 & Which backend, noise model, and mitigation metadata are associated with an execution? & Backend name, backend type, number of qubits, noise model, and error-mitigation method. \\
CQ5 & Which metadata are available for reproducing or auditing an execution? & Execution date, random seed, software package, and software version. \\
CQ6 & Are there executions marked as completed but missing result information? & Identifiers of completed executions without an associated result. \\
\bottomrule
\end{tabular}
\end{table*}

The questions cover the main information needed to understand, reproduce, compare, and debug \gls{vqa} executions. They retrieve the algorithmic components of an execution, the results of completed runs, the Pauli decomposition of the problem Hamiltonian, backend and noise-related metadata, software-environment information, and potentially incomplete execution records. In this role, \gls{sparql} complements \gls{shacl}. The validation layer checks whether a record is well formed, while the query layer retrieves the information needed to analyze the execution.

\begin{table}[t]
\centering
\small
\caption{Validation output for the example VQE execution records.}
\label{tab:validation-output}
\begin{tabular}{p{0.32\columnwidth}p{0.25\columnwidth}p{0.31\columnwidth}}
\toprule
\textbf{Validation aspect} & \textbf{Valid record} & \textbf{Invalid record} \\
\midrule
Overall conformance &
Conforms &
Does not conform \\

Number of violations &
0 &
20 \\

Execution result metadata &
Present and consistent &
Missing or incomplete result information \\

Hamiltonian representation &
Valid Pauli strings and coefficients &
Malformed Pauli terms and invalid values \\

Numerical constraints &
Ranges and datatypes satisfied &
Invalid numerical ranges and datatype errors \\

Controlled vocabularies &
Supported values used &
Unsupported vocabulary values detected \\
\bottomrule
\end{tabular}
\end{table}

\section{Evaluation and Discussion}

This section discusses the current evaluation of the proposed framework and the extent to which it supports representation, validation, and reproducibility-oriented inspection of \gls{vqa} execution records. The discussion is based on the implemented ontology, the \gls{vqe} case study, the \gls{shacl} validation results, and the \gls{sparql} competency queries.

\subsection{Coverage of \gls{vqa} Concepts}

The ontology covers the main entities required to describe \gls{vqa} executions, including algorithms, circuits, ansatzes, Hamiltonians, optimizers, backends, noise models, mitigation methods, software environments, execution steps, and results. The \gls{vqe} case study shows that these concepts are sufficient to represent a complete hybrid quantum--classical execution record, from the initial setup to the final numerical result. Although the evaluation focuses on \gls{vqe}, the ontology is not restricted to this algorithm. Its class structure also includes concepts relevant to other \gls{vqa} families, such as \gls{qaoa} and \gls{qml} workflows. The ontology deliberately focuses on execution metadata rather than low-level circuit semantics or hardware-control details.

\subsection{Validation and Query Results}

The validation and querying workflow was evaluated on \gls{vqe} execution records, including one valid record and one intentionally invalid record. he goal of this evaluation is not to benchmark large-scale \gls{rdf} storage
or validation performance, but to assess whether the proposed
representation can capture, validate, and query the execution metadata
needed for reproducibility in a controlled \gls{vqe} scenario.

The valid record conforms to the \gls{shacl} constraints, indicating that the required execution metadata are present and internally consistent. In contrast, the invalid record produces 20 validation violations corresponding to missing, malformed, or inconsistent metadata. These violations include missing result information, malformed Pauli terms, invalid numerical ranges, incorrect datatype values, and unsupported controlled-vocabulary values.

To support reproducibility and independent inspection, the replication
package contains the ontology, validation rules, example execution
records, competency queries, and scripts used in the case study.
Table~\ref{tab:replication-package} summarizes the main technical
artifacts included in the package.

\begin{table}[t]
\centering
\small
\caption{Summary of the technical artifacts in the replication package.}
\label{tab:replication-package}
\begin{tabular}{l r}
\toprule
\textbf{Artifact} & \textbf{Count} \\
\midrule
Ontology serializations & 2 \\
OWL classes & 71 \\
OWL object properties & 35 \\
OWL datatype properties & 33 \\
Ontology RDF triples & 528 \\
SHACL validation file & 1 \\
SHACL node shapes & 14 \\
SHACL property constraints & 33 \\
SHACL RDF triples & 269 \\
Example VQE execution records & 2 \\
SPARQL competency queries & 6 \\
Validation and query scripts & 2 \\
\bottomrule
\end{tabular}
\end{table}

The \gls{sparql} competency queries successfully retrieve the main execution components and reproducibility-relevant metadata from the valid record. In particular, they expose the ansatz, optimizer, backend, Hamiltonian terms, software information, final result, and execution metadata needed to inspect the run. The final competency question is consistency-oriented: for a complete valid record, it should return no result, whereas any returned execution indicates a run marked as completed without linked result information.

Together, the validation results and competency queries show that the proposed representation supports both automatic checking and query-based inspection of \gls{vqa} execution records. The validation output is summarized in \Cref{tab:validation-output}. The contrast between the valid and invalid records illustrates that the framework can distinguish complete execution records from records with reproducibility-relevant metadata defects. More broadly, the results show that semantic validation can identify problems that may not be visible from the final numerical result alone, supporting the automatic checking of execution records before they are reused, compared, or published.

\subsection{Usefulness for Reproducibility}

The framework makes the execution context of a \gls{vqa} run explicit and machine-readable. The \gls{sparql} competency queries retrieve information needed to reproduce or inspect an execution, including the ansatz, optimizer, backend, Hamiltonian terms, software versions, noise model, mitigation method, and final result. This helps distinguish the usual output of a \gls{vqa} run, such as final energy and optimized parameters, from the richer execution record needed for reproducible experimentation. By storing these details in a structured form, the framework can reduce manual inspection of source code, logs, configuration files, and paper descriptions.

\subsection{Limitations and Threats to Validity}

The main limitations concern scale, integration, and generalization. The current case study uses a small number of controlled \gls{vqe} records, while large benchmarking campaigns may contain thousands of executions, multiple backends, repeated runs, and many optimizer configurations. At that scale, additional work is needed to evaluate the performance of \gls{rdf} storage, \gls{shacl} validation, and \gls{sparql} querying over large execution datasets.

A second limitation is integration with existing quantum software frameworks. The ontology defines a structured representation for execution records, but practical adoption would require exporters from tools such as Qiskit \cite{javadi2024quantum}, PennyLane \cite{bergholm2018pennylane}, Cirq \cite{Cirq_Developers_2025}, or other workflow systems. Without such tooling, users would still need to construct \gls{rdf} records manually, which limits usability in real experimental pipelines.

A third limitation is ontology maintenance. Quantum software frameworks, backend interfaces, noise models, and mitigation techniques evolve quickly. The ontology therefore needs a versioning strategy and a process for adding new classes and properties without breaking existing execution records. This is especially important for long-term reproducibility, where old records should remain interpretable even as software and hardware platforms change.

Finally, the validation rules reflect the metadata requirements selected in this work. Other research groups may require different levels of detail, for example more hardware calibration information, more detailed optimizer traces, or framework-specific execution options. Future work should therefore evaluate the framework on larger datasets, additional \gls{vqa} families, and execution records generated automatically from different quantum software environments. The case study therefore demonstrates feasibility rather than scalability.

\section{Conclusion}

This paper presented an ontology-supported framework for representing \gls{vqa} execution records as structured and machine-readable software-engineering artifacts. The proposed ontology captures not only the usual numerical output of a \gls{vqa} run, such as the final energy and optimized parameters, but also the execution context needed to reproduce, validate, compare, and reuse the result. In particular, the ontology models key entities involved in \gls{vqa} experimentation, including algorithms, ansatzes, Hamiltonians, optimizers, backends, noise models, mitigation methods, execution steps, software environments, measurement outcomes, and results.

The framework combines this semantic representation with \gls{shacl} validation and \gls{sparql} querying. The validation layer makes metadata requirements explicit and detects missing, malformed, or inconsistent execution information. The query layer supports reproducibility-oriented inspection by retrieving information about execution components, completed runs, Hamiltonian terms, backend and noise settings, software versions, and incomplete records. The \gls{vqe} case study demonstrates how valid execution records can be represented and inspected, and how intentionally invalid records can be identified automatically.  The representation is also applicable to broader hybrid quantum--classical workflows in which quantum execution is coupled with classical optimization, software orchestration, and \gls{hpc} or cluster-based computational resources.

Overall, the results suggest that ontology-based representations can support more reliable \gls{vqa} experimentation by making execution records explicit, checkable, and reusable. This is relevant for quantum software engineering because \gls{vqa} results depend on many interacting software, hardware, numerical, and configuration choices that are often scattered across code, logs, configuration files, and paper descriptions. Representing this information in a structured form can reduce manual inspection and improve the comparability of experimental artifacts.

Future work will focus on extending the evaluation beyond controlled \gls{vqe} records. This includes applying the ontology to larger benchmarking datasets, additional \gls{vqa} families such as \gls{qaoa} and \gls{qml} workflows, and records generated from different quantum software frameworks. Another direction is the development of tool support for automatically exporting execution records from existing quantum software environments, which would make the proposed representation easier to adopt in practical experimental pipelines.

\section*{Data Availability}

The ontology, \gls{shacl} validation shapes, \gls{sparql} queries, example \gls{vqa} execution records, and validation outputs are made publicly available through a Zenodo repository \cite{illesova_2026_vqa_execution_ontology}. The repository also includes valid and intentionally invalid \gls{vqe} example records, the validation scripts, query files, and documentation needed to reproduce the validation and querying results reported in this paper.

\bibliographystyle{ACM-Reference-Format}
\bibliography{references}

\end{document}